# TeV/m Nano-Accelerator: Investigation on Feasibility of CNT-Channeling Acceleration at Fermilab

Y. M. Shin[1, 2], A. H. Lumpkin[2], and R. M. Thurman-Keup[2]

[1]*Department of Physics, Northern Illinois University, Dekalb, IL, 60115, USA*

[2]*Fermi National Accelerator Laboratory (FNAL), Batavia, IL, 60510, USA*

**ABSTRACT**

The development of high gradient acceleration and tight phase-space control of high power beams is a key element for future lepton and hadron colliders since the increasing demands for higher energy and luminosity significantly raise costs of modern HEP facilities. Atomic channels in crystals are known to consist of 10 – 100 V/Å potential barriers capable of guiding and collimating a high energy beam providing continuously focused acceleration with exceptionally high gradients (TeV/m). However, channels in natural crystals are only angstrom-size and physically vulnerable to high energy interactions, which has prevented crystals from being applied to high power accelerators. Carbon-based nano-crystals such as carbon-nanotubes (CNTs) and graphenes have a large degree of dimensional flexibility and thermo-mechanical strength, which could be suitable for channeling acceleration of MW beams. Nano-channels of the synthetic crystals can accept a few orders of magnitude larger phase-space volume of channeled particles with much higher thermal tolerance than natural crystals. This paper presents the current status of CNT-channeling acceleration research at the Advanced Superconducting Test Accelerator (ASTA) in Fermilab.

## 1. INTRODUCTION

A great interest of the scientific community has focused on the idea of building a post-Large Hadron Collider (LHC) machine, such as the International Linear Collider (ILC), Compact Linear Collider (CLIC), or Future Circular Collider (FCC). This is based on the belief that these machines can provide sufficient center-of-mass (CM) energy and luminosity a few orders of magnitude higher than the current levels for the new physics experiments [1 – 3]. The prospective milestone of the super-colliders is envisioned to achieve ~ 100 TeV in CM energy and/or ~ $10^{34}$ - $10^{35}$ $cm^{-2}s^{-1}$ in luminosity. The biggest challenge for this grand plan is to achieve the CM energy and luminosity within a realistic facility footprint and with the maximum possible power conversion from the "wall-plug" to the beams. Therefore, a realistic machine may have to fit within a fairly limited footprint, which might be about 10 kilometer or less, and with less than a few tens of MW beam power [4]. Will such colliders be capable of reaching energies that may be orders of magnitude larger than current ones, namely, 100 – 1000 TeV? It is quite questionable that with current acceleration technology the facility can actually reach the particle energy and luminosity needed for new physics experiments within the size and power constraints. To obtain the energies of interest within the given footprint, development of ultra-fast acceleration with tight phase-space control of high power beams will be the most important mission for future lepton and hadron colliders and it will only be possible with multidisciplinary and multilateral approaches.

## 2. CRYSTAL CHANNELING – TeV/m ACCELERATION

Shock waves in an ionized plasma media, excited by relativistic particles, have been of great interest on account of their promise to offer extremely high acceleration gradients of $G$ (max. gradient) = $m_e c \omega_p / e \approx 96 \times n_0^{1/2}$ [V/m], where $\omega_p = (4\pi n_p e^2/m_e)^{1/2}$ is the electron plasma frequency and $n_p$ is the ambient plasma density of [cm$^{-3}$], $m_e$ and $e$ are the electron mass and charge, respectively, and $c$ is the speed of light in vacuum. However, a practically obtainable plasma density ($n_p$) in ionized gas is limited to below ~ $10^{24}$ m$^{-3}$, which in principle corresponds to wakefields up to ~ 100 GV/m [5 – 7], and it is realistically very difficult to create a stable gas plasma with a charge density beyond this limit. Metallic crystals are the naturally existing dense plasma media completely full with a large number of conduction electrons available for the wakefield interactions. The density of charge carriers (conduction electrons) in solids $n_0 = $ ~ $10^{26} – 10^{29}$ m$^{-3}$ is significantly higher than what was considered above in gaseous plasma, and correspondingly the wakefield strength of conduction electrons in solids, if excited, can possibly reach 10 TV/m in principle.

Figure 1 shows our simulation graphs of a beam-driven wakefield acceleration in a homogeneous plasma model with solid-level charge densities, $10^{25}$ m$^{-3}$ and $1.6 \times 10^{28}$ m$^{-3}$, which might be in the lower and upper limits of electron density of solid-state media (acceleration gradient and energy gain versus beam charge density) [8]. The densities are considered for the assessment as the corresponding plasma wavelengths, 10 μm and 0.264 μm, are also in the spectral range of available beam-modulating sources such as IR/UV lasers or magnetic undulators (inverse FELs) [9] for strong beam-plasma coupling in the beam-driven acceleration. As shown in the figure, with the beam-driven acceleration the acceleration gradient ranges from 0.1 TeV/m to 10 TeV/m with the solid plasma densities at the linear beam-plasma coupling condition ($n_b \sim n_p$). The gradient corresponds to 10 – 30 MeV of energy gain with a channel length of 10 plasma wavelengths. Also, it appears that the beam is strongly focused and collimated by the large transverse fields generated from the oscillating plasma over the distance (Figs. 1(c) and (d)). Apparently, this effective plasma simulation model shows TeV/m range gradient with solid state level plasma density. However, in spite of the exceptionally large amplitude of accelerating and focusing fields, in reality if charged particles are injected into either an amorphous solid or a random orientation of a single phase crystal, they will encounter a large stopping power from nuclear and electron scatterings. Also, irregular head-on collisions at atomic sites completely spread the particles all over, accompanied with phase-space volume expansion. Particles will thus be quickly repelled from a driving field due to fast pitch-angle diffusion of large scattering rates/angles from non-uniformly distributed atoms.

The channels between atomic lattices or lattice planes aligned in a crystal orientation of natural crystals like silicon or germanium are sparse spaces with relatively low electron densities [10]. Charged particles, injected into a crystal orientation of a mono-crystalline (homogeneous and isotropic) target material, undergo much lower nucleus and electron scatterings. The channeled particles can be accelerated in two different ways, depending upon how they gain energies from driving sources (lasers or particle-beams) through plasma wakefields of conduction electrons, or diffracted EM fields confined in the crystal channels. Plasma wakefields in crystals, if excited by a strong driving source, either a short particle-bunch or a high power laser, can accelerate particles along the space in the lattice channel. For the channeling acceleration with confined x-ray diffraction, the atomic channel can hold > $10^{13}$ V/cm transverse and $10^9$ V/cm longitudinal fields of diffracted traveling EM-waves from an x-ray laser coupled to a crystal at the Bragg diffraction angle ($\lambda/2a = \sin\theta_B$), where $a$ is the lattice constant and $\theta_B$ is the diffraction angle). However, to accelerate channeled particles with high gradients by confined x-ray fields, the acceleration requires coherent hard x-rays ($\hbar\omega \approx 40$ keV) of power density $\geq 3 \times 10^{19}$ W/cm$^2$ to overcome radiation losses of channeled muons [11], which exceed those conceivable today. The x-ray pumping method thus fits for heavy particles, e.g. muons and protons, which have relatively smaller radiation losses. For electrons (and positrons), the beam-driven acceleration is more favorably applicable to channeling acceleration as the energy losses of a drive beam can be transformed into the acceleration energy of a witness beam [12]. The maximum particle energies, $E_{\max} \approx (M_b/M_p)^2 (\Lambda G)^{1/2} \{G/(z^3 \times 100 \text{ [GV/cm]})\}^{1/2} 10^5$ [TeV], can reach 0.3 TeV for electrons/positrons, $10^4$ TeV for muons, and $10^6$ TeV for protons ($M_b$ is the mass of the channeled particle, $M_p$ is the proton rest mass, $\Lambda$ is the de-channeling length per unit energy, and $z$ is the charge of the channeled particle) [13]. Here, channeling radiation of betatron oscillations between atomic planes is the major source of energy dissipation. The de-channeling length ($\Lambda$) is also a critical factor, in particular in the low energy regime, since it scales as $E^{1/2}$ [14]. The idea of accelerating charged particles in solids along major crystallographic directions was suggested by several scientists such as Pisen Chen, Robert Noble, Richard Carrigan, and Toshiki Tajima in the 1980's and 1990's [15 – 19] for the possible advantage that periodically aligned electrostatic potentials in crystal lattices are capable of providing a channeling effect [20 – 26] in combination with low emittance determined by an Ångström-scale aperture of the atomic "tubes". The basic concepts of atomic accelerator with short pulse driving sources like high power lasers or ultra-short bunches have been considered theoretically. However, the idea has never been demonstrated by experiment or simulation due to the extremely tight interaction condition of the Angstrom-size atomic channels in natural crystals and the complexity of electron dynamics in solid-plasma.

## 3. CHANNELING ACCELERATION IN CARBON NANOTUBES

Carbon nanotubes (CNTs) are a synthetic nanostructure, which is a roll of a graphene sheet, and its tube diameter can be easily increased up to sub-micron by optimizing fabrication processes (chemical vapor deposition, CVD). For channeling applications of high power beams, carbon nanostructures have various advantages over crystals [27]. Particles are normally de-channeled when the transverse forces are larger than the maximal electric field acting on channeled particles from crystal atoms, which is described by the critical angle ($\varepsilon_{cn}$ (normalized rms acceptance) = $(1/2)\gamma a \theta_c$, where $\theta_c$ is the critical angle). The dechanneling rate is significantly reduced and the beam acceptance is dramatically increased by the large size of the channels, e.g., a 100 nm wide CNT channel has larger acceptance than a silicon channel by three orders of magnitude. Some previous studies [28] on the radiative interaction in a continuous focusing channel present an efficient method to damp the transverse emittance of the beam without diluting the longitudinal phase space significantly. CNTs can efficiently cool channeled particles similar to natural crystals. The strong focusing in a CNT channel results in a small beta function, so that it is quite feasible to create a beam of small transverse emittance.

If the channel size is increased from angstroms to nanometers (Figs. 2(a) and (b)), the maximally reachable acceleration gradient would be lowered from ~ 100 TeV/m to ~ 1 TeV/m due to the decrease of effective plasma charge density. However, the nanotube channels still provide sufficiently large transverse and longitudinal fields in the range of TV/m. For the crystal channels in angstrom scale, the lattice dissociation time of atomic structures ( $\Delta t \approx \sqrt{(m_i/m_e)}(2\pi/\omega_p)$ ), where $m_i$ and $m_e$ are the masses of ion (carbon) and electron respectively [29]) is in the range of sub-100 fs with 1 TV/m fields, corresponding to $10^{19}$ W/cm$^3$. For beam driven acceleration, a bunch length with a sufficient charge density would need to be in the range of the plasma wavelength to properly excite plasma wakefields, and channeled particle acceleration with the wakefields must occur before the ions in the lattices move beyond the restoring threshold and the atomic structure is fully destroyed. It is extremely difficult to compress a particle bunch within a time scale of femto-seconds since the bunch charge required for plasma wave excitation and the beam power corresponding to the time scale will exceed the damage threshold of the crystal. The disassociation time is, however, noticeably extended, to the order of pico-seconds, by increasing the channel size to nanometers because the effective plasma density and corresponding plasma frequency are decreased by a few orders of magnitude, as shown in Fig. 2(a). The constraint on the required bunch length is thus significantly mitigated (Fig. 2(c)) and the level of power required for an external driving source could be lowered by a few orders of magnitude, although the acceleration gradient will be lowered accordingly [30].

Furthermore, dephasing length [31] is appreciably increased with the larger channel, which enables channeled particles to gain a sufficient amount of energy (Fig. 2(d)). The atomic channels in natural crystals, even if they provide extremely high potential gradients, are limited to angstroms and are unchangeable due to the fixed lattice constants. The fixed atomic spaces make the channeling acceleration parameters impractically demanding, but CNTs could relax the constraints to more realistic regimes. The carbon structures comprised entirely of covalent bonds (*sp*2) are extremely stable and thermally and mechanically stronger than crystals, steel, or even diamonds (*sp*3 bond). It is known that thermal conductivity of CNTs is about 20 times higher than that of natural crystals (e.g. silicon) and the melting point of a freestanding single-tip CNT is 3,000 – 4,000 Kelvin. Carbon based channels thus have significantly improved physical tolerance against intensive thermal and mechanical impacts from high power beams.

## 4. SIMULATION ANALYSIS WITH FERMILAB-ASTA 50 MEV BEAM

The initial assessment of the prospective energy gains of CNT-channeled electrons was already fulfilled with a PIC-based beam-driven plasma simulator combined with the beamline simulations. For the simulations, the ASTA 50 MeV beamline from the chicane (BC1) to the imaging station (X124) was modeled with CST and Elegant (Fig. 3(a)). The two beam profiles of the bunches with and without the modulation (modulation wavelength = $\lambda_{mb}$), which is generated by the slit-mask in the bunch compressor (BC1), are monitored at the goniometer position. The beam profiles are then manually imported to the effective CNT model [32]. A typical simulation result (Fig. 3(b)) showed that a 1 nC bunch (uncorrelated energy spread = ~ 0.01 – 0.015 %, $\lambda_{mb}$ = 100 μm) is self-accelerated with a net energy gain (~ 0.2 %) on the tail (witness) and an energy loss (~ 0.6 %) on the head (drive) along the 100 μm long channel with the nominal beam parameters. Our preliminary assessment with the full beamline model predicts that the ASTA 50 MeV beam can produce ~ 1 – 2 % of maximum net gain with 3.2 nC bunch charge and 100 μm transverse beam size (circular beam, Fig. 3 (c)), corresponding to 5 – 10 GV/m gradient. For the simulations, the bunch charge density is about a thousand times smaller than the channel charge density (off-resonance beam-plasma coupling). However, detecting the amount of energy gain by the proposed experiment will support feasibility of TeV/m acceleration in CNT channels.

Generally, density modulations enhance energy efficiencies or power gains of coherent light sources or beam-driven accelerators. A beam, if longitudinally modulated, is more strongly coupled with accelerating or undulating structures at a resonance condition with the fundamental or higher order modes. Pre-bunched or modulated beams would improve the longitudinal beam control in energy-phase space and furthermore strongly enhance wakefield strengths or transformer ratio of beam-driven channeling accelerations. The modulated beam either maximizes the wakefield strength (when it is bunched with a plasma wavelength of an accelerating medium) or significantly increases the transformer ratio proportionally to the number of micro-bunches ($R = M \cdot R_0$, where $M$ is the number of micro-bunches and $R_0$ is the transformer ratio of a single-bunch driver) with an off-resonance condition [33]. In principle, a beam-density modulation (or micro-bunching) corresponding to an intrinsic channel plasma frequency possibly offers the optimum beam-plasma coupling condition with maximum energy transfer efficiency and acceleration gradient.

However, the crystal charge density with $\rho_p \geq 10^{25}$ m$^{-3}$ requires a very short modulation wavelength, $\lambda_{mb} \leq 10$ μm, which could be only produced by a short-period micro-buncher (e.g. Inverse FEL undulator). Our plan for the experiment is thus to generate a bunched beam of relatively long modulation periodicity by slit-masking the beam in the center of a magnetic chicane and then to couple a higher order mode of the modulated beam with an accelerating medium [34]. The slit-mask modulation technique is relatively easy to generate a beam modulation. At the ASTA low energy beamline, while a ~ 3 – 4 mm long photo-electron bunch passes through the bunch compressor (BC1), the slit-mask placed in the BC1 will slice bunches into micro-bunch trains by imprinting the shadow of a periodic mask onto the bunch with a correlated energy spread (Fig. 4(a)). In principle, modulation strength and periodicity of the modulation can be controlled by adjusting the grid period or by the dipole magnetic field [35]. The bunch-to-bunch distance (modulation periodicity) is given as,

$$\Delta z = W \frac{\sqrt{(1+h_1 R_{56})^2 \sigma_{z,i}^2 + \tau^2 R_{56}^2 \sigma_{\delta i}^2}}{\eta_{x,mask} h_1 . \sigma_{z,i}} \approx W \frac{|\sigma_{z,i} + R_{56}\sigma_\delta|}{\eta_{x,mask} \sigma_\delta},$$

($\sigma_{z,i}$: initial bunch length, $h_1$: first order chirp = $- 1/R_{56}$, $\sigma_{\delta,i}$: initial uncorrelated energy spread, $R_{56}$: longitudinal dispersion of BC1, and $\tau$: energy ratio = $E_{io}/E_{fo}$, $E_{io}$ and $E_{fo}$ are the central energies before and after acceleration, respectively, and $\eta_{x,mask}$: dispersion at the mask). With nominal ASTA beam parameters ($\sigma_{z,i}$ = 3 ps, $R_{56}$ = ~ - 0.192 m, $E_{io}$ = 50 MeV), the analytic model showed that a slit-mask with slit period 900 μm and aperture width 300 μm generates ~ 100 μm spaced micro-bunches with 2.4% correlated energy spread. As shown in Fig. 4(c), the preliminary simulation data from Elegant and CST also indicated that the designed mask produces a ~ 100 μm spaced beam modulation with maximum RF chirp.

## 5. EXPERIMENTAL PERSPECTIVES

It is planned to test the CNT-channeling acceleration at the Fermilab ASTA 50 MeV beamline (Fig. 5(a)). A ~ 3 ps long electron bunch generated from the photo-injector is transported to the magnetic chicane (BC1). The bunch is compressed to ~ 1 ps by BC1, and the compressed beam will be focused by the quadrupole triplet magnets (Q118, 119, 120). After the beam spot size is focused to ~ 100 – 200 μm, it will be injected to a CNT target in the goniometer. The CNT targets will be fabricated by an anodic aluminum oxide (AAO)-CNT template process. The process is a well known nanofabrication technique to implant straight, vertical CNTs in a porous aluminum oxide template by the chemical vapor deposition (CVD) growth process [36 – 40]. Sub-100 μm long straight multi-wall CNTs grow along the aligned nano-pores in an AAO template, which is followed by thermal cleaning. AAO templates with 100 μm long, 20 – 200 nm wide pores are commercially available [41, 42]. The channeling test of the AAO-CNT with $^4$He+ beam was already successful in a low energy regime (2 MeV) [43]. The channeled electrons will be transported to an electron spectrometer consisting of a dipole (D122) and a screen in the energy dispersive region following D122 at imaging station X124. Their energy distribution will be measured by the spectrometer before the beam is dumped to a shielded concrete-enclosure (beam dump).

The pre-bunched beam generated by the slit-mask installed at X115 between two bending-dipoles (D115 and D116) will also be tested and the measured beam parameters will be compared with the ones of the bunched beam to check the impact of beam-modulation on channeling acceleration. Before testing the AAO-CNT target, the beam parameters will be characterized first without a target, which will be a reference for the beam-energy measurement. An experiment will be set up to check if the measured variation of the projected image on the screen due to presence of the target exceeds the nominal deviation of the image produced by the intrinsic energy spread. The initial experiment will then be followed by subsequent measurements to accurately identify net energy gains/losses and beam emittances: the channeled beam deflected by the magnetic spectrometer (D122) will be projected on the screen of the imaging station (X124). The beam-injection angle with respect to the target axis will then be scanned. A relative change of projected images from one angle to another will be translated into an energy gain/loss of channeled beam (Fig. 5(b)).

## 6. CONCLUSION

Particle channeling in crystalline media has been considered an alternative promising technology for challenging scientific experiments at the high energy physics. Its strong atomic interactions along the crystal lattice planes enable efficient collimation/bending of intense beams and continuously focused acceleration with exceptionally large gradients up to a few tens of TeV/m. Development of the channeling acceleration concept has a great potential to advance accelerator technology for future HEP colliders.

**ACKNOWLEDGMENT**

This work was supported by the DOE contract No. DEAC02-07CH11359 to the Fermi Research Alliance LLC. We thank Vladimir D. Shiltsev of Accelerator Physics Center (APC) in Fermi National Accelerator Laboratory (FNAL) for the helpful discussion on the idea and the support for the experimental plan.

# References

[1] https://www.linearcollider.org/ILC

[2] http://clic-study.web.cern.ch/CLIC-Study/

[3] https://espace2013.cern.ch/fcc/Pages/default.aspx

[4] V. D. Shiltsev, Physics – Uspekhi 55 (10) 965 (2012)

[5] T. Tajima and J. M. Dawson, Phys. Rev. Lett. 43(4), 267 (1979).

[6] J. B. Rosenzweig, D. B. Cline, B. Cole, H. Figueroa, W. Gai, R. Konecny, J. Norem, P. Schoessow, and J. Simpsion, Phys. Rev. Lett. 61, 98 (1988).

[7] C. Joshi and T. Katsouleas, Phys. Today 56(6), 47 (2003).

[8] Y. Shin, "Beam-driven acceleration in ultra-dense plasma media", Appl. Phys. Lett. 105, 114106 (2014)

[9] I. Y. Dodin and N. J. Fisch, Phys. Plasmas 15, 103105 (2008).

[10] M. A. Kumakhov and F. F. Komarov, Energy Loss and Ion Ranges in Solids, Gordon and Breach Science (1979)

[11] T. Tajima, and M. Cavenago, PRL 59, 1440 (1987)

[12] P. Chen and R. J. Noble, AIP. Conf. Proc. 156, 2122 (1987)

[13] B. W. Montague and W. Schnell, in: Laser Acceleration of Particles, eds. C. Joshi and T. Katsouleas, AIP Conf. Proc. No. 130 (1985)

[14] P. Chen and R. Noble, "Channel Particle Acceleration By Plasma Waves in Metals", slac-pub-4187

[15] L. A. Gevorgyan, K. A. Ispiryan, and R. K. Ispiryan, JETP Lett. 66, 322 (1997).

[16] P. Chen and R. J. Noble, AIP Conf. Proc. 156, 222 (1987).

[17] P. Chen and R. J. Noble, in *Relativistic Channeling*, edited by R. A. Carrigan and J. Ellison Plenum, New York, 1987, p. 517; also NATO ASI Ser., Ser. B 165, 517 1987; SLAC-PUB-4187 1987.

[18] P. Chen and R. J. Noble, AIP Conf. Proc. 396, 95 1997; also FERMILAB-CONF-97-097 1997; SLAC-PUB-7673 1997.

[19] F. Zimmermann and D. H. Whittum, Int. J. Mod. Phys. A 13, 2525 1998; also SLAC-PUB-7741 1998.

[20] D. S. Gemmell, Rev. Mod. Phys. 46, 129 (1974).

[21] J. Lindhard, Mat.-Fys. Medd. Dan. Vid. Selsk., Vol. 34, No. 14 (1965); also in Usp. Fiz. Nauk 99, 249 (1969).

[22] V. V. Beloshitsky, F. F. Komarov, and M. A. Kumakhov, Phys. Rep. 139, 293 (1986).

[23] V. M. Biryukov, Yu. A. Chesnokov, and V. I. Kotov, *Crystal Channelingand Its Application at High-energy Accelerators* (Springer, New York, 1997).

[24] V. N. Baier, V. M. Katkov, and V. M. Strakhovenko, *Electromagnetic Processes at High Energies in Oriented Single Crystals* (World Scientific, Singapore, 1998).

[25] R. A. Carrigan Jr., J. Freudenberger, S. Fritzler, H. Genz, A. Richter, A. Ushakov, A. Zilges, and J. P. F. Sellschop, Phys. Rev. A 68, 062901 (2003)

[26] B. Newberger, T. Tajima, F.R. Huson, W. Mackay, B.C. Covington, J.R. Payne, Z.G. Zou, N.K. Mahale, S. Ohnuma, Proc. IEEE Part. Acc. (IEEE, Chicago, 1989), p. 630.

[27] M. Murakami, and T. Tanaka, APL 102, 163101 (2013)

[28] Z. Huang, P. Chen and R. D. Ruth. Phys. Rev. Len. 74 (10) (1995) 1759.

[29] P. Chen and R. J. Noble, SLAC-PUB-7402 (1998)

[30] Young-Min Shin, Dean A. Still, and Vladimir Shiltsev, "X-ray driven channeling acceleration in crystals and carbon nanotubes", Phys. Plasmas 20, 123106 (2013)

[31] C. B. Schroeder, C. Benedetti, E. Esarey, F. J. Grüner, and W. P. Leemans, Growth and Phase Velocity of Self-Modulated Beam-Driven Plasma Waves, Phys. Rev. Lett. 107, 145002 (2011)

[32] http://www.txcorp.com/home/vsim/vsim-pa

[33] E. Kallos, *Plasma Wakefield Accelerators using Multiple Electron Bunches*, PhD Dissertation of Univ. Southern California (Electrical Engineering) 2008

[34] D. C. Nguyen, B. E. Carlston, "Amplified coherent emission from electron beams prebunched in a masked chicane", Nuclear Instrument and Methods in Physics Research A 375, 597 (1996)

[35] P. Muggli, V. Yakimenko, M. Babzien, E. Kallos, and K. P. Kusche, "Generation of Trains of Electron Microbunches with Adjustable Subpicosecond Spacing", PRL 101, 054801 (2008)

[36] Tariq Altalhi, Milena Ginic-Markovic, Ninghui Han, Stephen Clarke, and Dusan Losic, *Membranes 1*(1), 37 – 47 (2011)

[37] A. Larsen, " Nano-Scale Convective Heat Transfer of Vertically Aligned Carbon Nanotube Arrays", Project Number: MQP-JNL-CNT9

[38] H. Peng Xiang, L. Chang, S. Chao, and C. HuiMing, Chinese Science Bulletin 57, 187 (2012)

[39] T. Altalhi, M. Ginic-Markovic, N. Han, S. Clarke, D. Losic, "Synthesis of Carbon Nanotube (CNT) Composite Membranes", Membranes 2011, 1, 37-47;

doi:10.3390/membranes1010037

[40] P. Ciambelli, L. Arurault, M. Sarno, S. Fontorbes, C. Leone, L. Datas, D. Sannino, P. Lenormand, B. Du Plouy Sle, "Controlled growth of CNT in meso-porous AAO through optimized conditions for membrane preparation and CVD Operation", Nanotechnology 22 (26), 265613 (2011)

[41] http://www.sigmaaldrich.com/labware/labware-products.html?TablePage=109501875

[42] http://www.synkerainc.com/

[43] Zhiyuan Zhu, Dezhang Zhu, Rongrong Lu, Zijian Xu, Wei Zhang, Huihao Xia, "The experimental progress in studying of channeling of charged particles along nanostructure", International Conference on Charged and Neutral Particles Channeling Phenomena, Proc. of SPIE Vol. 5974 (SPIE, Bellingham, WA, 2005)

**Figure Captions**

FIG. 1 Acceleration gradient versus normalized charge density graphs (top: (a) and (c)) and beam/plasma charge distributions (bottom: (b) and (d)) of multi-bunched beam with (a) and (b) $n_p = 10^{25}$ m$^{-3}$ and (c) and (d) $n_p = 1.6 \times 10^{28}$ m$^{-3}$

FIG. 2 (a) Effective plasma density, (b) acceleration gradient, (c) dissociation time scale, and (d) dephasing length versus channel size (carbon-based).

FIG. 3 (a) ASTA beamline model (CST). Inset is a modulated bunch charge distribution (Elegant and CST) at the goniometer position and effective CNT-channeling acceleration model (VORPAL). (b) Energy distributions of a modulated bunch (top) without and (bottom) with a channel. (c) energy gain versus bunch charge

FIG. 4 (a) Conceptual drawing and (b) simulation model of a slit-mask micro-buncher (c) longitudinal charge distributions and beam signal spectra obtained by Elegant and CST simulations

FIG. 5 (a) ASTA 50 MeV beamline configuration (b) schematics for energy and emittance measurements of CNT-channeling experiment (dotted green box in (a))

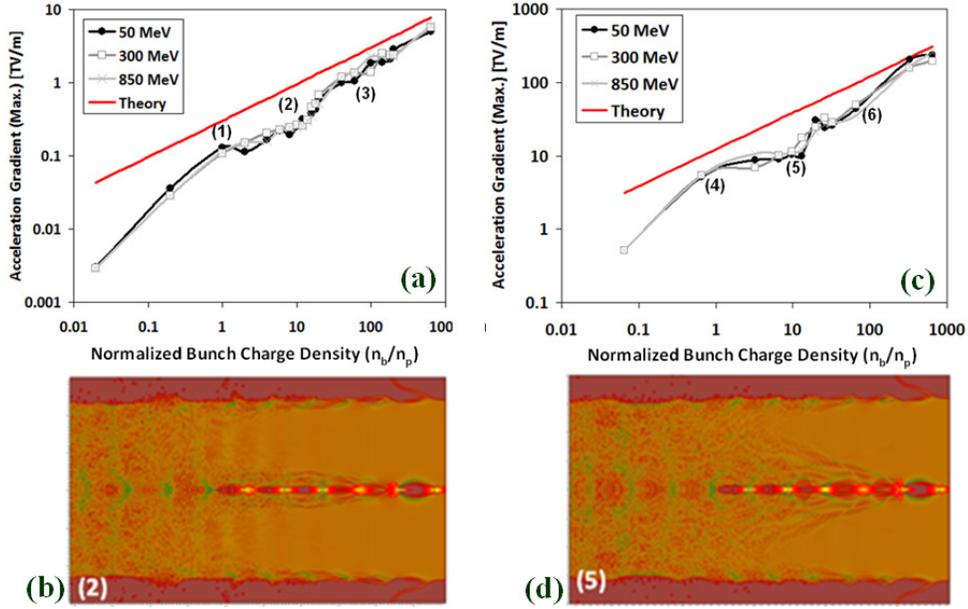

FIG. 1 (Y. M. Shin)

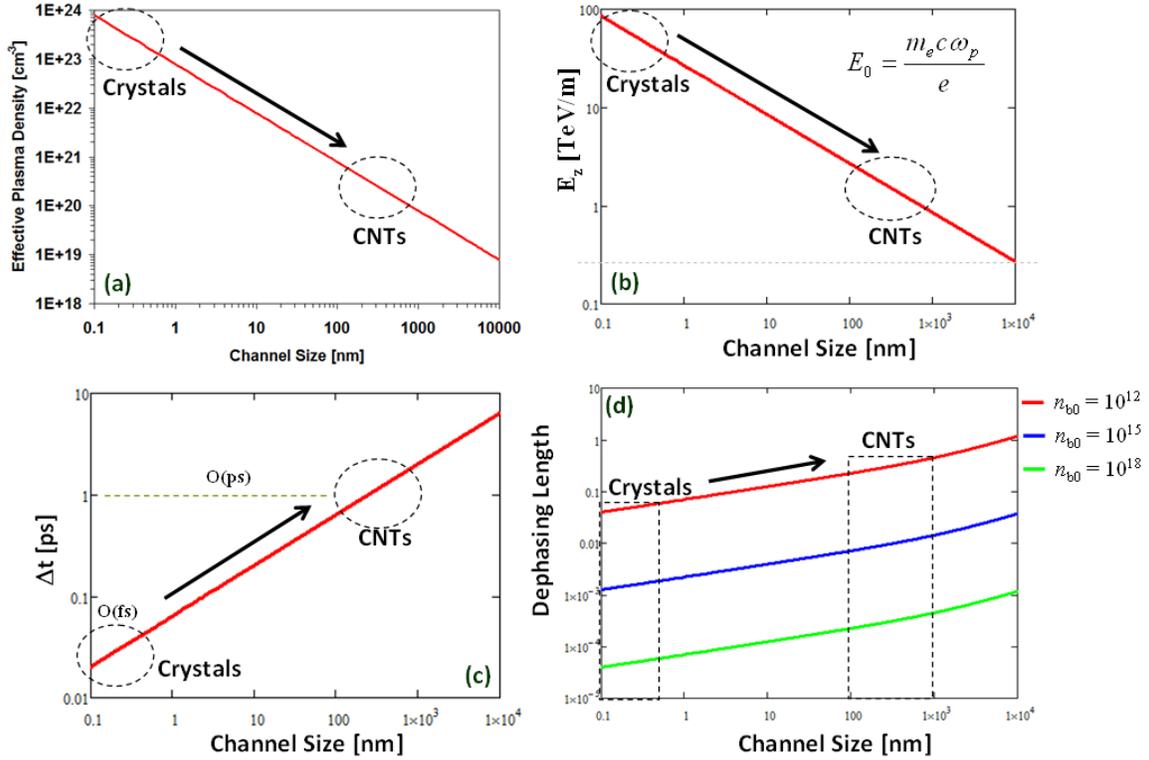

FIG. 2 (Y. M. Shin)

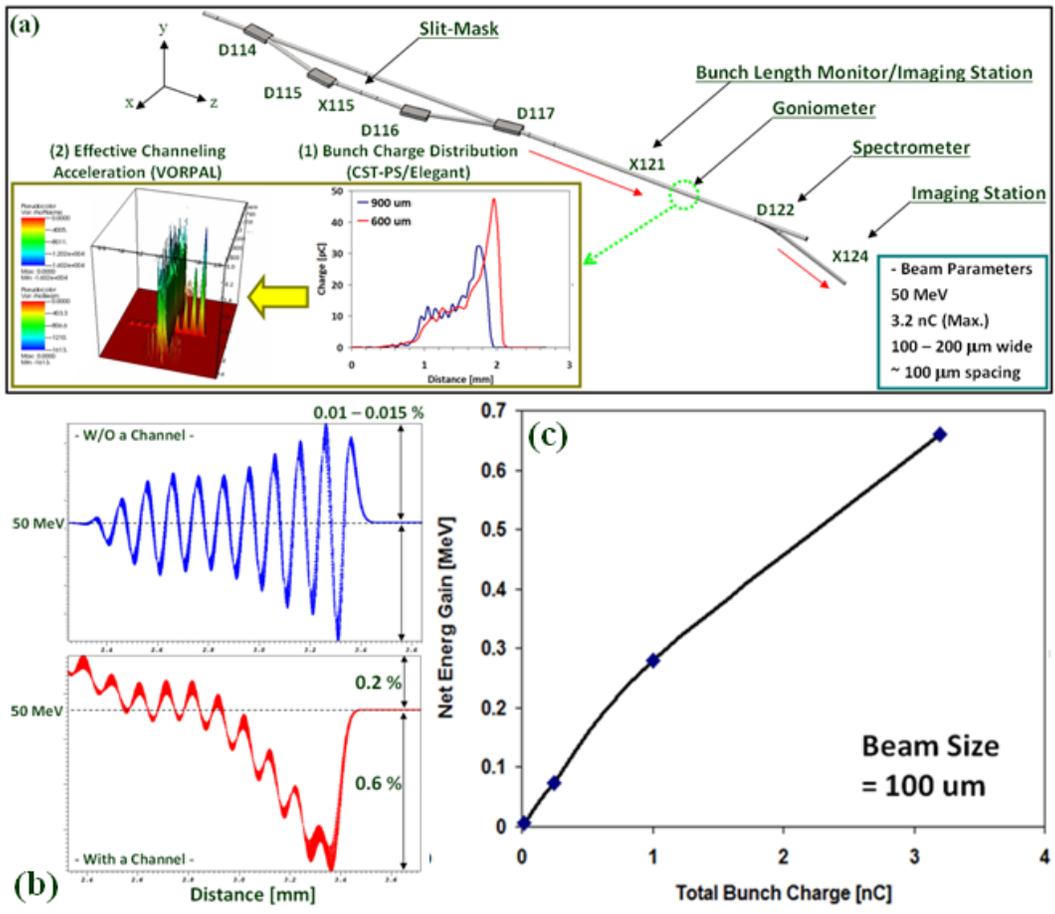

FIG. 3 (Y. M. Shin)

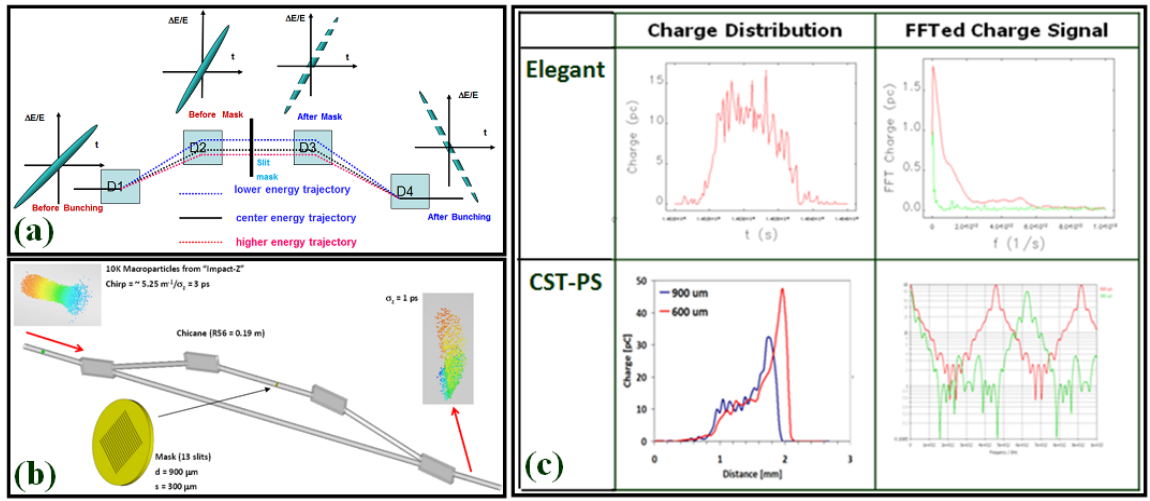

FIG. 4 (Y. M. Shin)

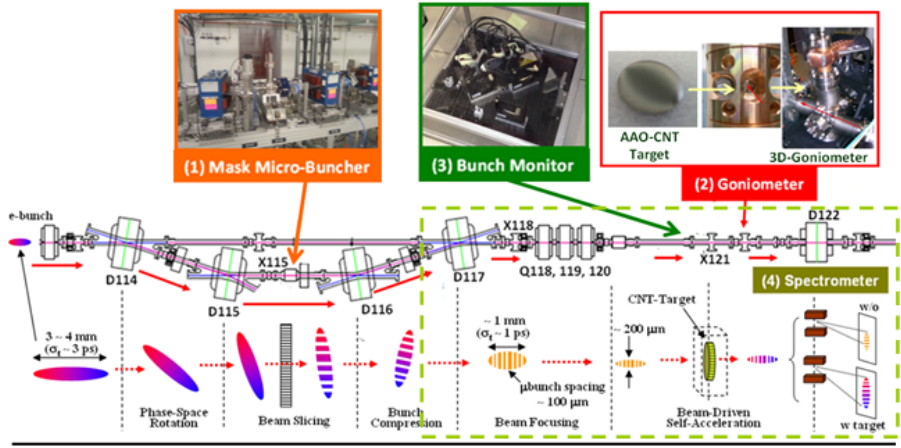
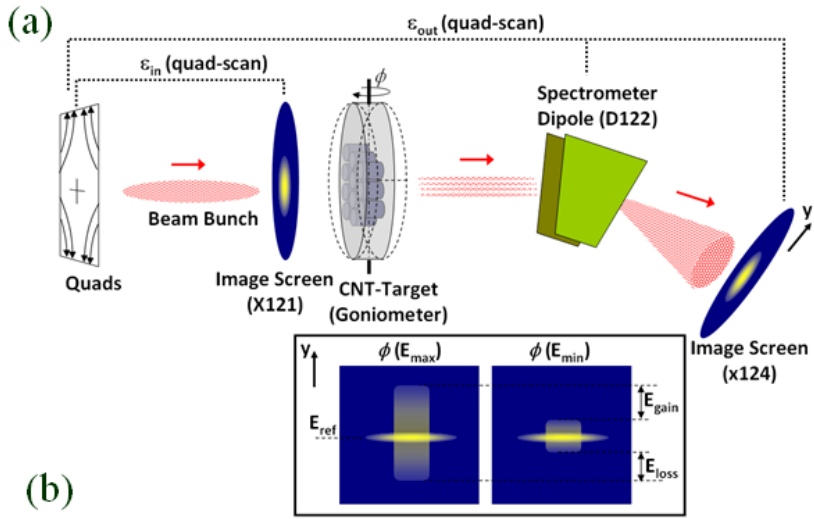

FIG. 5 (Y. M Shin)